\documentclass[conference]{IEEEtran}
\IEEEoverridecommandlockouts
\usepackage{cite}
\usepackage{multirow}
\usepackage{tabu}
\usepackage{tabularx}
\usepackage[pdftex]{graphicx}
\usepackage[cmex10]{amsmath}
\usepackage{amssymb}
\usepackage{algorithmic}
\usepackage{array}
\usepackage{mdwmath}
\usepackage{eqparbox}
\usepackage{caption}
\usepackage{subcaption}
\usepackage{url}
\usepackage{booktabs}
\usepackage{color}
\usepackage{tikz-network}
\usepackage{bm}
\usepackage[linesnumbered,ruled]{algorithm2e}
\usepackage{flushend}
\usepackage{csquotes}

\newcommand{\printfnsymbol}[1]{%
  \textsuperscript{\@fnsymbol{#1}}%
}
\usepackage[bookmarks=false]{hyperref}

\newcommand{\eg}{\emph{e.g}. }
\newcommand{\ie}{\emph{i.e}. }

\begin{document}

\title{Resolving the cybersecurity Data Sharing Paradox to scale up cybersecurity via a co-production approach towards data sharing}

\author{
\IEEEauthorblockN{Amir Atapour-Abarghouei\IEEEauthorrefmark{1}\textsuperscript{\textsection},
A. Stephen McGough\IEEEauthorrefmark{1}\textsuperscript{\textsection} and
David S. Wall\IEEEauthorrefmark{2}\textsuperscript{\textsection}
}

\IEEEauthorblockA{\IEEEauthorrefmark{1}School of Computing, Newcastle University \\ \{Amir.Atapour-Abarghouei, Stephen.McGough\}@newcastle.ac.uk}
\IEEEauthorblockA{\IEEEauthorrefmark{2}Centre for Criminal Justice Studies, University of Leeds\\ D.S.Wall@leeds.ac.uk}
}
\maketitle

\begingroup\renewcommand\thefootnote{\textsection}
\footnotetext{All authors made equal contribution to this work.}
\endgroup

\begin{abstract}

  As cybercriminals scale up their operations to increase their profits or inflict greater harm, we argue that there is an equal need to respond to their threats by scaling up cybersecurity. To achieve this goal, we have to develop a co-productive approach towards data collection and sharing by overcoming the cybersecurity data sharing paradox. This is where we all agree on the definition of the problem and end goal (improving cybersecurity and getting rid of cybercrime), but we disagree about how to achieve it and fail to work together efficiently. At the core of this paradox is the observation that public interests differ from private interests. As a result, industry and law enforcement take different approaches to the cybersecurity problem as they seek to resolve incidents in their own interests, which manifests in different data sharing practices between both and also other interested parties, such as cybersecurity researchers. The big question we ask is can these interests be reconciled to develop an interdisciplinary approach towards co-operation and sharing data. In essence, all three will have to co-own the problem in order to co-produce a solution. We argue that a few operational models with good practices exist that provide guides to a possible solution, especially multiple third-party ownership organisations which consolidate, anonymise and analyse data. To take this forward, we suggest the practical solution of organising co-productive data collection on a sectoral basis, but acknowledge that common standards for data collection will also have to be developed and agreed upon. We propose an initial set of best practices for building collaborations and sharing data and argue that these best practices need to be developed and standardised in order to mitigate the paradox.

\end{abstract}

\begin{IEEEkeywords}
    Cybersecurity, Data Sharing, Cyber Attacks, Big Data, Artificial Intelligence
\end{IEEEkeywords}

\section{Introduction}
\label{sec:intro}

Cybercrime is becoming an all too familiar feature of the world we live in these days and it seems that a day does not go by without another major breach of an important online system. In May 2020, for example, cloud computing provider, Blackbaud, was attacked, allegedly by the AKO ransomware group, using a form of ransomware (ransomware 2.0) which exfiltrates the victim's key data and the data of its many hundreds of clients and their millions of customers \cite{gatlan2020blackbaud} and published the fact (with evidence) on the attacker’s leak web site. Once stolen, the data is used to leverage a ransom payment, or is sold/auctioned off to the highest bidder~\cite{eduhack}. Other forms of ransomware, for example, Wannacry, brought the UK National Health Services to a standstill \cite{crunch}.

In a recent twist, attackers are now targeting companies and organisations who host big data. They infiltrate the organisation, exfiltrate their key data before encrypting it in order to lever a ransom~\cite{wall2020double}. Cybersecurity and law enforcement seek to thwart these attacks and keep organisations safe. However, the sharing of security datasets which can be used by the cybersecurity community to improve their response is, at best, somewhat lacking.

At the heart of the cybersecurity mission is an agreement as to what the central problem is, for example, cybercrime. But also an agreement as to what the end goal is, for example to eradicate cybercrime by prevention, mitigation and prosecution. A common problem experienced across the cybersecurity sector, however, is that whilst everyone agrees about the problem, there is considerable disagreement about how to achieve the goal. Public interests greatly differ from the private interests. For example, policing agencies want to work with victims, investigate the offence and prosecute the offenders in the public interest. Industry and many other organisational victims, on the other hand, simply want to resolve the issue, not just restoring their systems to what they were before the attack, but also not alerting competitors, the public and, importantly, their shareholders to the fact that they have been victimised. The third group here are the cybersecurity researchers whose primary interest is to collect and analyse data from cyber-incidents.  Here, we define cybersecurity researchers as individuals or groups who seek to analyse and learn from the data collected from a cyber-incident in order to provide findings that change the current practices and reduce the chances of future breaches, irrespective of whether these researchers come from industry, academia or policing. 

This paradoxical difference is usefully illustrated in ransomware cases, which provide a stark example to show this paradox. On one side are the police agencies who seek more open reporting and cooperation to investigate victimisations and pursue the offenders. The victims, on the other hand, along with the cyber-insurance and third-party cybersecurity companies discretely employed by them not only tend to pay the ransom as standard practice, but also seek to negotiate with the offenders to reduce the ransom demands, all against public policy. The cybersecurity researchers, however, seek to collect and analyse data from these ransomware attacks in order to predict and prevent future incidents. This is where their objectives fail to synchronise with the aims of the other two groups.

Cybersecurity researchers seek access to real-world (big) data, which can be used to develop new techniques for identifying and blocking attacks (one rare example of this type of data is the SheildFS dataset for ransomware attacks~\cite{continella2016shieldfs}). Though obtaining these datasets is often a hard process, thwarted with the dangers of organisations not wishing to divulge that they have been under attack as outlined above. Likewise, interactions between researchers and the public sector policing agencies tend to be less than fruitful as the policing agencies often do not own the data themselves nor do they see data for future prevention and detection as a key benefit for themselves.

So, three key players - private sector, public sector policing agencies and cybersecurity researchers - take very different approaches to the cybersecurity problem as they seek to resolve it for their own particular interests. This contradiction is at the heart of what we refer to as the “cybersecurity data sharing paradox” and may explain why it is hard, if not impossible, for the different sectors to work together without intervention.

\begin{figure}[t!]
	\centering
	\includegraphics[width=0.99\linewidth]{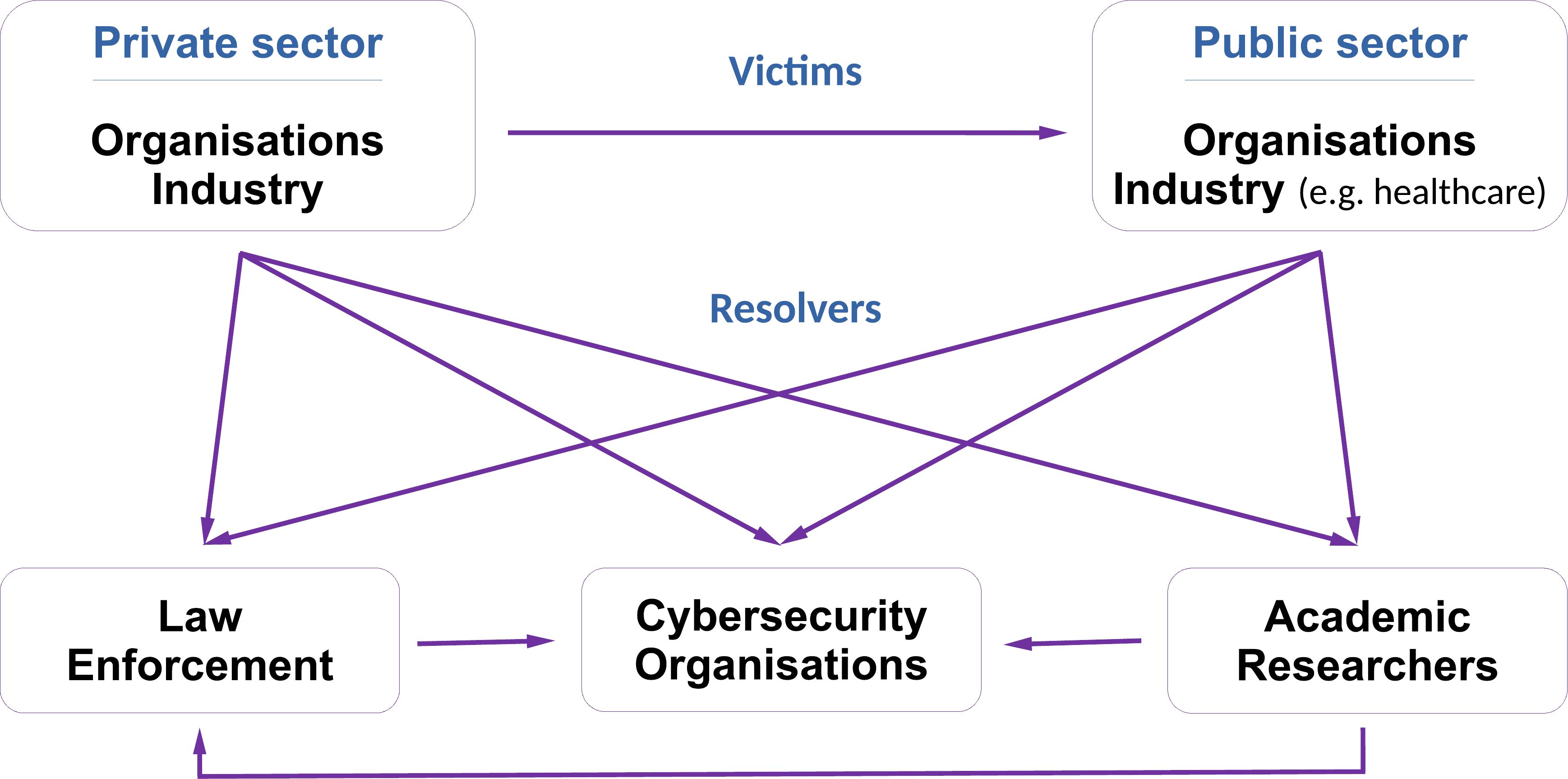}
	\captionsetup[figure]{skip=7pt}
	\captionof{figure}{\textbf{In an Ideal World} - The many relationships in the Cybersecurity Data Sharing Paradox.}%
	\label{fig:relationships}\vspace{-0.5cm}
\end{figure}

Complicating this paradox are the many additional dimensions of sub-interests created by the different relationships in the equation: industry and law enforcement, industry and cybersecurity industry, industry and academia, academia and cybersecurity industry, academia and law enforcement. Furthermore, within each sub-sector are also disciplinary divisions which can sometimes compete or have different orientations or obligations, for example, private and public sector organisations, or in law enforcement, local and national police, or in academia, social and computational sciences. In an ideal world, the many different relationships are expressed as outlined in Figure \ref{fig:relationships}. In this figure, the key relationships between the different players are outlined. Lines indicate the potential interactions which we would argue should exist between the different players. In many cases, however, these interactions do not exist, or if they do, they are far less effective than they should be. 

The question we ask in this paper is \enquote{can these interests be reconciled to develop an interdisciplinary approach towards co-operation and sharing data?} In essence, all three have to co-own the problem to co-produce a solution; a phrase that is easy to say, but hard to achieve, but we argue that a few operational models with good practices do exist that provide a possible solution, especially multiple third-party ownership of organisations (\eg UK Payments\footnote{https://www.ukpayments.org.uk/what-we-do/} - formerly APACS - Association for Payment Clearing Services) which aggregate and analyse their sector's payment clearing data. To take this forward, perhaps the data could be organised on a sectoral basis, for example, as per those sectors listed in Q5 of the National Data Strategy Policy Paper by the UK Department for Digital, Culture, Media \& Sport \cite{DCMS}. In this case, agreements on standards for data collection will have to be reached. If a third-party approach is not adopted, then this will lead to one-to-one relationships needing to be formed, which are all too often slow to develop and fail to scale. In this paper, we seek to outline a set of issues to shape a future discussion about developing standards, procedures and best practices in general around data collection. Discussion of these issues will help to add granularity upon implementation of the proposed UK National Data Strategy \cite{DCMS} which, at the time of writing, was still out for consultation. 

The rest of this paper is set out as follows. In Section \ref{sec:motivation}, we provide exemplar cases to illustrate the nature of the problem at hand and to motivate the need for addressing the paradox. We present existing datasets in Section \ref{sec:datasets} and argue that these are neither sufficient in quantity nor comprehensive enough. We reflect and analyse the 2019 workshop we held at the Alan Turing Institute on data challenges in Section \ref{sec:workshop} before discussing how we can move things forwards in Section \ref{sec:forward}. Finally, we conclude this paper in Section \ref{sec:conclusion}.

\section{The Recent Increase in the Scalability of Cybercrimes}
\label{sec:motivation}

\begin{figure}[t!]
	\centering
	\includegraphics[width=0.99\linewidth]{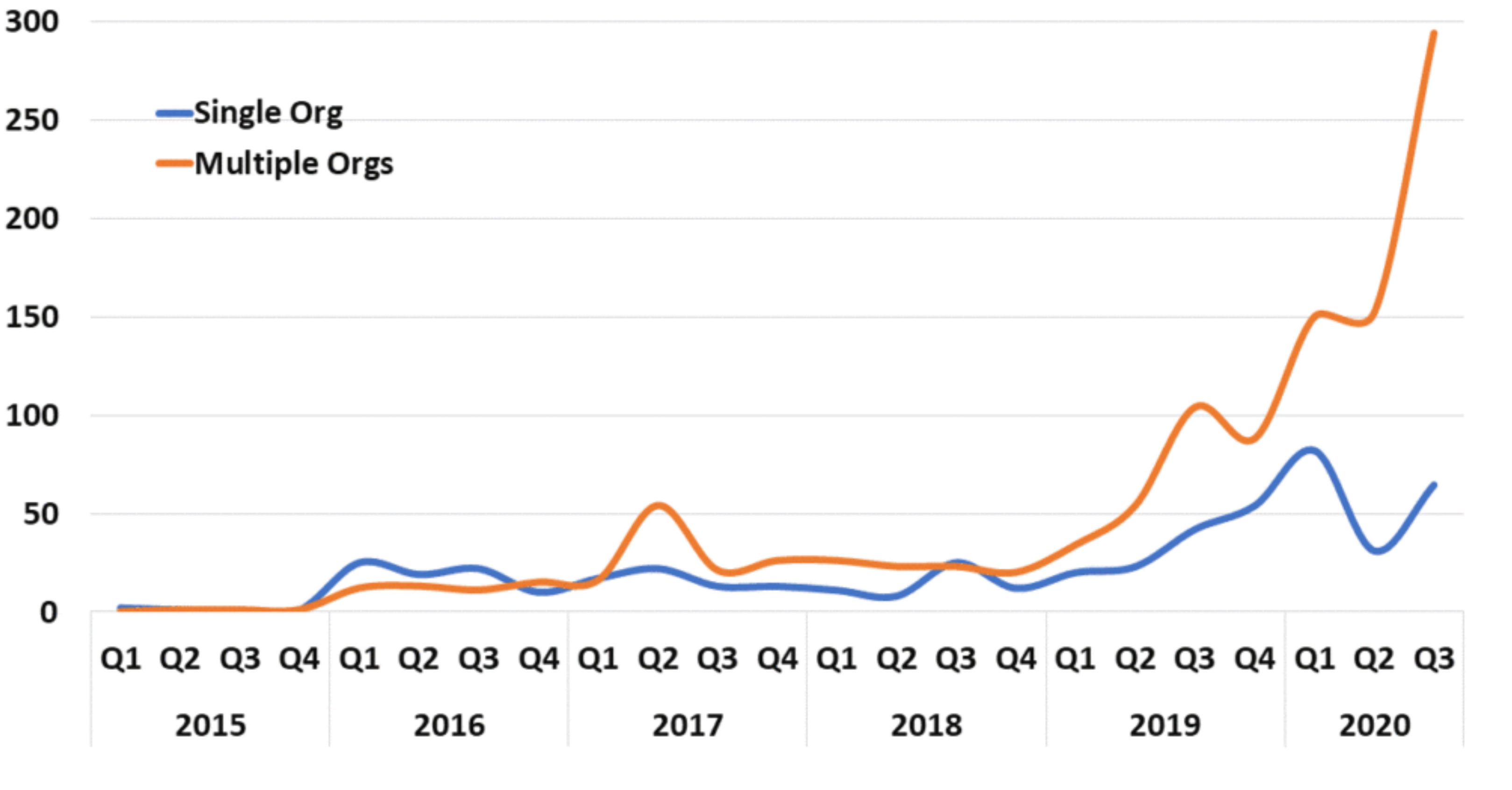}
	\captionsetup[figure]{skip=7pt}
	\captionof{figure}{Single vs. multiple (complex) organisational victims \cite{wall2020double}. The number of cases analysed in this figure is 2000.}%
	\label{fig:graph}\vspace{-0.5cm}
\end{figure}

During recent times, there has been a change in cybercrime attack vectors, which has increased both the scalability of cybercrime and also the harms to society. This is best expressed by changes in ransomware. Figure \ref{fig:graph} shows how attacks on larger organisations (multiples) have, since 2019, dramatically scaled up their impact down the supply chain by focusing upon multiple cloud service providers \cite{wall2020double}. They not only directly affect their clients, but also their client’s clients. We conservatively estimate that each attack impacts upon about 15 client organisations and in some cases many more.

Most specifically, new forms of blended ransomware attacks (ransomware 2.0 \cite{vanhorn}) now include the added fear tactic of ‘naming and shaming’ (or reverse double jeopardy bis in idem) \cite{wall2020double}. This is in contrast to the previous generation of ransomware, which relied on ‘spray and pray’ tactics that tempted millions of recipients with juicy subject lines in the expectation that some of them would reply or open attachments that would infect their computer or start an infection process \cite{connolly2019rise}. The new generation of attacks are the result of careful research and planning by criminals and the targeting of senior managers to get their access \cite{wall2020hackers}. The new generation is therefore strategically different from the old one. Using stolen (or bought) login details, attackers enter the victim’s computing system and copy key organisational data before encrypting it. Reports suggest that attackers could have been within the system for a year (or more) to prepare the ground for the attack. In the past year, they have also adopted a new tactic of publishing the victim’s name on a website that they control along with some proof of attack. By publicly ‘naming and shaming’ victims, attackers can leverage the extortion of the ransom payment. Furthermore, attackers very often demand a ransom (sometimes in the millions), which if not paid (in Bitcoin) within a set time period (such as 7 days) is doubled and more data is published. Some ransomware gangs ask for two ransoms, one for the encryption code to make the system work again and another to delete the sensitive data stolen. They may also, even, be prepared to negotiate down the final ransom amount to match the victim’s budget.

Not only are complex organisations now direct primary targets for attackers, but their outsourced service clients also become secondary victims when they are attacked. In May 2020, cloud computing provider, Blackbaud, was attacked by ransomware. Blackbaud is a cloud technology company used by the educational sector (\eg  schools, colleges \& universities) and also various not-for-profit organisations and the healthcare sector. Its many (possibly thousands) clients worldwide became secondary victims when their data, stored by Blackbaud, was potentially compromised. Ransomware now deliberately seeks to steal or deny the victim the use of their data as a ransom leverage tactic, hence, ransomware should arguably be additionally classed as a data crime. Blackbaud is a useful case study to explain the increase in scalability of attacks \cite{gatlan2020blackbaud} and also the data problem that arises. Not least the consequences of such attacks, because Blackbaud are now the focus of various class actions being brought by victims \cite{gatlan2020blackbaud}. 

Having explained the increase in scalability of cyberattacks, most notably in the context of ransomware, it is important to consider what data should be collected and shared to reduce the number and success of these attacks? What does the data look like? What data needs to be shared and what does not? Indeed, ethically what data can be shared? How can data relating to breaches be shared with impunity? These issues will be discussed later, next we will explore the issues of motivation and also the tactics being used by offenders which will need to be countered.

\subsection{Understanding and Defining the Problem}

Ransomware attacks, as stated earlier, involve data theft, which is amongst other offences, a key-stone crime. Once data is stolen, it not only has a ‘ransom’ value to the owner, but it also has a secondary value to others who can process the stolen data. Big data offenders use Artificial Intelligence (AI) algorithms to identify sub-groups of potential victims, for example, lawyers, teachers, health workers or managers and tailor phishing attacks to them. Or they might simply use any credentials within the data to gain access to their systems and their employer’s systems. We have referred to this ‘criminal data flow’ as the cascade effect \cite{porcedda2019cascade}. Big data, of whatever sort, helps fuel big crimes \cite{wall2018big}. In addition to utilising big data to victimise, offenders are also using (artificially) intelligent malware to, for example, seek out vulnerable systems, obfuscate their malware, enter the systems, and also obscure their activities once in.

Offenders use AI to increase their advantage over their victims. But if criminals are using data and AI to attack, should not cybersecurity be doing the same to defend? However, we need to separate out offensive and defensive AI cybersecurity systems as these require different approaches. We also have to separate out preventative, mitigating and investigating systems as they seek separate sub-goals within the larger cybersecurity mission.

But, the question remains, how do cybersecurity researchers get the data they need, what does it look like and how do we address the ‘garbage in, garbage out’ problem to avoid bias? Can we, for example, learn lessons from IBM DeepLocker \cite{stoecklin2018deeplocker}, the AI Cybercrime Simulator? IBM developed Deeplocker to \enquote{conceal the malicious intent in benign unsuspicious looking applications, and only triggers the malicious behavior once it reaches a very specific target, who uses an AI model to conceal the information, and then derive a key to decide when and how to unlock the malicious behavior} \cite{patterson2018weaponized}. Hackers with artificial intelligence are problematic for law enforcement, because it helps them increase their scalability by keeping one step ahead, especially if the AI can decide for them which computer can be attacked most effectively. However, the class of malware indicated by Deeplocker has not yet been experienced \cite{stoecklin2018deeplocker}, but the question is not if but when, so there is still time to prepare a response. And evidence from ransomware development and evolution is suggesting that ransomware can evaluate the ‘worth’ of the victim to calculate the most appropriate ransom.

Within the EMPHASIS Ransomware research project Atapour-Abarghouei et al. \cite{atapour2019king} used AI to identify ransomware types from screen images of the ransomware note. In another project, AI systems were developed to help identify illegal data exfiltration \cite{mcgough2015detecting, mcgough2015insider}. Basically, the message from the ‘grey’ cybersecurity literature is that criminal use of AI is evolving and that we have to learn from their cybercrime playbook and apply AI routines to key parts of the cybercrime ecosystem to respond to attacks.

\section{What is the issue with data and sharing it?}
\label{sec:datasets}

At the heart of the problem being addressed is the need to collect and share data. Whilst there are many datasets available, they are often created for different purposes to cyber security, or they lack common standards in data collection. Cybercrime statistics, for example, can rarely be compared because data about economic cybercrimes are not usually compatible with cyber-pornographic images, or hacking/computer misuse, or cyber-bullying or social media harassment. Data needs to be captured with appropriate metadata, such as what attack was going on and what mitigations were being used at the time. Likewise, it is of little value to collect data only when an attack is taking place as this will lack comparative data for ‘normal’ situations. Hence the need for best practice in collection, aggregation and analysis.

The ever-increasing number of data breaches and security attacks observed on a regular basis\cite{atapour2019volenti} and the innovative use of novel attacks by cybercriminals emphasise the importance of getting ahead of the curve using cutting-edge techniques such as AI. Modern artificial intelligence and machine learning approaches \cite{atapour2019monocular, simonyan2014very, atapour2019veritatem, ren2015faster, atapour2019complete, bonner2019temporal, atapour2020rank, maciel2019online, atapour2019generative, al2020beyond} have revolutionised numerous areas of research with significant industrial, societal and economic impact. Making use of such AI-based methods to predict and prevent breaches and attacks would give the cybersecurity industry the advantage they urgently need. A significant challenge in developing AI techniques, however, is the need for neatly curated accurately-labelled data which, as explained previously, is extremely rare and not easily shared when it comes to security breaches.

As discussed earlier, ransomware is an excellent representative of the modern cybercrime paradigm, as it is capable of victimising highly targeted organisations and individuals along with any indiscriminate home user and can inflict irreversible harm on its victims. The “No More Ransom” project~\cite{nomoreransom} provides a mechanism to identify the ransomware from either the text within the ransomware note or a small number of the encrypted files. Using a large database of information on previously identified ransomware variants, this project is specifically dedicated to helping all victims, whether individual home users or businesses, to recover their encrypted files without having to pay the ransom to the perpetrators.

ShieldFS \cite{continella2016shieldfs}, an add-on driver that works towards protecting the Windows native file system from ransomware attacks, provides a large-scale dataset of I/O request packets (IRP) generated by benign applications as well as active ransomware samples. The data includes about 1.7billion IRPs produced by 2,245 different applications running on clean machines and systems infected with the main ransomware families (\ie CryptoWall, TeslaCrypt, Critroni, CryptoDefense, Crowti). The dataset is large, varied and highly generalisable, but it is broadly captured, and certain fine-grained features and labels are missing, making it impractical for certain machine learning techniques.	

Atapour-Abarghouei et al. \cite{atapour2019king} provides a dataset of ransom notes and splash screens often displayed on systems infected with various forms of ransomware. The dataset includes the splash screens and ransom notes of 50 different variants of ransomware. A single image of a splash screen variant is available for each of the ransomware classes available with some classes associated with more than one splash screen (\ie certain classes contain more than one training image, but those images depict different splash screens associated with the same class). The dataset provides a balanced test set of 500 images (10 images per class) to evaluate any ransomware identification techniques.

As for security breaches in companies and organisations, the data is often withheld from the public, making any form of data analysis and machine learning training more difficult. The US Office of Civil Rights data breach portal provides an online database describing data breaches of protected health information (PHI) that affect 500 or more individuals \cite{hhs}, \cite{gabriel2018data}. Minimal data is provided in terms of the type of breach (\eg hacking/IT incident, improper disposal, loss, theft, unauthorised access/disclosure) and the location or mode of the breached information (\eg desktop computer, electronic health records, email, laptop computer, network server, paper/films). While this is an excellent source of data for geographic and demographic analysis of vulnerabilities in healthcare data, due to the limitations of the features available in the dataset, it cannot be used for AI or even any detailed and concrete conclusions about the causes and effects of such data breaches.	

In another somewhat similar dataset \cite{mccandless2020world}, the world’s biggest breaches are regularly recorded (and visualised) with features including the company /organisation breached, the type of company / organisation, type of breach, data sensitivity, news references and a description of the events surrounding the breach. Despite being a very useful source of data, technical details of the data breach are not clear, and the dataset cannot therefore be used as a source of training data for a machine learning system.

With origins in Verizon’s Data Breach Investigations Reports, VERIS (Vocabulary for Event Recording and Incident Sharing) \cite{veris} is now widely established in the cyber-security community and aims to encourage a spirit of collaborative data sharing with respect to cyber-security incidents by helping organisations to collect useful information and share them with the research community. Additionally, VERIS offers a set of metrics and common terminology for describing security incidents in a structured and organised manner for better analysis and reproducibility \cite{moreira2018extending}. VERIS structures itself around indicators common to all cyber-security incidents, including the Action used to breach the asset, the Actor who breached the asset, the compromised Asset, the security Attribute (confidentiality, integrity or availability) that was affected \cite{oosthoek2020hodl}.

VERIS comprises two primary elements: the schema\footnote{http://github.com/vz-risk/veris} and the dataset\footnote{https://github.com/vz-risk/VCDB}. The dataset consists of a collection of incidents documented in individual files identified by Universally Unique Identifiers. While the dataset contains data from a large number (more than 8,000) of incidents, the primary issue with the project is that the majority of data is provided by the team responsible for the project and a small number of partner organisations. Despite the excellent collaborative opportunities this framework offers for data sharing, the existing data is limited in detail and quality and only a fraction of the records fully utilise all the fields provided for technical details. VERIS epitomises the challenges of data sharing and further emphasises the importance of resolving the cybersecurity data sharing paradox.

Not only are the number of publicly available datasets low, but the quality of these datasets are also insufficient for serious analysis and AI. This is compounded by the fact that these datasets go quickly out of date as cybercriminals are constantly evolving their approaches.

In order to take the data collection, compilation and sharing issue forwards we now draw upon the outcomes of a Turing workshop on Machine Learning and data challenges.

\section{Building on the ``Machine Learning and data challenges'' Alan Turing workshop} 
\label{sec:workshop}

In June 2019 we held a workshop at the Alan Turing Institute in London\footnote{https://www.turing.ac.uk/events/machine-learning-and-data-challenges-ransomware-and-cloud} on Machine Learning and data challenges in ransomware and the cloud. This event was to kick-start the process of identifying best practices in big data collection of cybersecurity incidents. The event was attended by more than 40 people from across industry, academia, government, law enforcement and the third sector. Researchers were from computing, cybersecurity and the social sciences. During this workshop we raised a number of questions to groups of attendees. The questions are listed in the following.

\begin{itemize}
	\item How can the problem be co-owned and the solution be co-produced?
	\item What sort of language is used to express the problem of accessing data?
	\item How can researcher expectations of data providers be managed and vice versa?
	\item Although researchers and data providers ultimately have the same goals and how do they view each other?
	\item How can the sharing of data be encouraged? - What would `good' look like?
	\item What is best practice to encourage data sharing?
	\item How should data be anonymised?
	\item What security requirements should be in place for data sharing? - What core principles need to be established?
\end{itemize}

Below we summarise the outcomes of these group sessions. In what follows we define two types of entities within the problem domain - those of data providers (or the owners of the data) and data consumers - those who wish to analyse the data. In general data producers are from industry, though there were examples of government, law enforcement, third sector and even academics being data providers. Data consumers tend to be from academia apart from those who work in the cybersecurity industry. As the terms data providers and consumers provide a clearer way of distinguishing between the attendees of the event we will use these descriptions from here on. It should be noted that a data provider may not be the data producer, however, as our interest here is in who owns the data we do not elaborate further on this distinction.

\begin{figure*}[t!]
	\centering
	\includegraphics[width=0.99\linewidth]{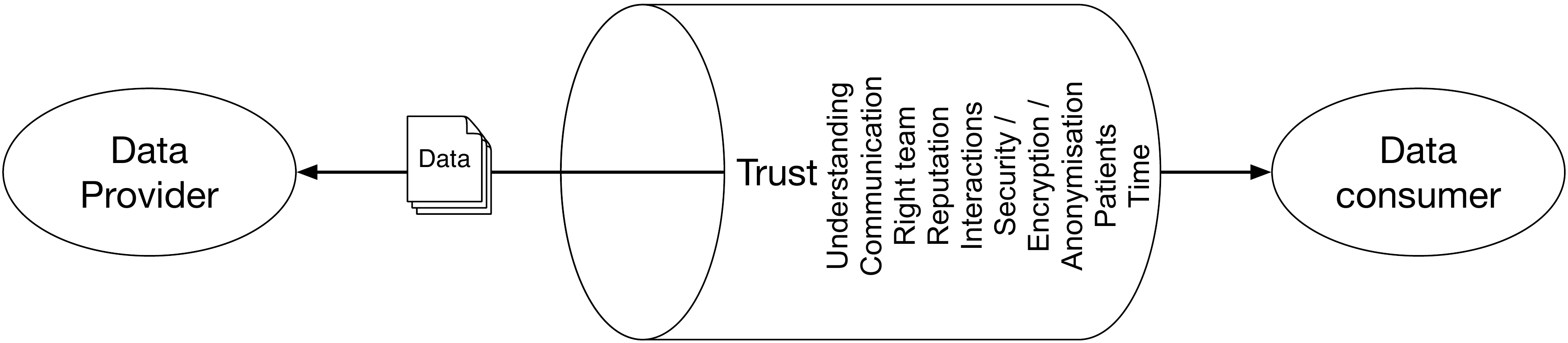}
	\captionsetup[figure]{skip=7pt}
	\captionof{figure}{Building up a trust conduit to enable sharing.}%
	\label{fig:trust}\vspace{-0.5cm}
\end{figure*}

In order for data providers to share with data consumers a level of trust needs be established between the two parties. This can be illustrated as in Figure \ref{fig:trust}. Here we depict trust as a pipeline between the two parties. However, the pipeline is fragile and requires a number of constructs to be established and nurtured, these include understanding, communication, the right set of people, building reputation, constant interaction, a full adherence to best security practices (including encryption and anonymisation) as well as an appreciation of time -- where one party may be slow at some times but other times wish to go fast. If any of these constructs fail the pipeline will collapse and sharing will not be possible. We discuss these constructs further in the rest of this section.

\subsection{Why may data providers not interact with academia?}

There was a perceived level of cynicism towards academia with many potential data providers feeling that academia has naive, arrogant and simplistic views of the problem and how they can influence the outcome. Academics were seen as ‘chasing funding’ where a funder would put out a call, researchers would then chase the call and try to fit what they had to the call rather than look for the right solution for the particular data provider. Academics were seen as often more interested in novel work than work fit for purpose. Then moving on once they have achieved their goal and not providing long-term solutions. 

Data providers are naturally, and understandably, wary about sharing their data - this is especially true after an attack incident, which is, unfortunately, often when academics will approach them. There was a perceived fear over reputational damage -- through exposure of their own internal bad practices. There was also the perception that academics did not have a clear plan of what they wished to do with the data -- stemming from the often taken viewpoint in academia of `give us the data and we can then let you know what we can do with it’. A further highlighted perception was that providing the data to an academic could lay the provider open to other non-intended risks such as legal challenges -- supposing that the academic discovered that the data was in breach of some legal requirement. There was also a perception that academics ignored the ethical issues within the problem domain -- something the providers could not ignore. Most of the issues discussed can be resolved via the concept of trust, which many data providers felt did not yet exist. Finally, participants highlighted the fact that \enquote{it’s easier to say ‘no’}. Yet, despite the mis-perceptions, the incentives to share data with others (including academics) were clearer and simpler, not least, improving the public good, financial incentives both for the provider and others, and the addition of better security by preventing the attacks being successful in the future.

\subsection{Problems with time}

Both providers and consumers cited issues over timings. Data consumers complained that it would take too long to get hold of the data whilst providers complained that it took too long for the consumers to come back with results from the data that they had shared. The feeling was that these issues could be rectified, if not at least reduced, if both parties had a clearer understanding of what each had to do.

\subsection{How to build trust}

Trust came over as one of the key concepts which needed to be correct within any data provider -- data consumer relationship. With trust being required in both directions. Honesty was seen as one of the primary requirements here along with clear and concise communication. Lack of understanding of what both parties could provide and wished to receive lead to misunderstandings and hence a loss of trust. This requires a deep understanding of each other’s values, problems, expectations and objectives. Though it was appreciated that for both parties these may change over the course of a collaboration -- however, as long as this is communicated then it can be handled. The feeling was that this trust could not be created instantaneously and would require long term interactions with regular meetings -- and building on human interactions to foster trust. A number of people cited that trust was best developed if the work was co-created and both parties gained positive outcomes. Both parties stated a value in minimising surprises within the process. This can be achieved through managing expectations, admitting failures and seeking regular feedback. 

It was felt by most that it was a bad idea to go for a significant interaction at the outset, but much better to start small with low-risk examples which were easy for both parties to work on and had fairly rapid in turn-around. This not only allowed the different parties to better understand each other but fostered greater trust as achievements were being made. This does require both parties to be more flexible. 

The development of a formal agreement between data provider and data consumer was seen by most as a best practice. This could take the form of a memorandum of understanding, full contractual agreement, or a non-disclosure agreement. The scale and level of this agreement would depend on the nature of the work undertaken and the perceived risk. These documents should define such concepts as who owns what (including the original data, derived data and any IP which may come out of the work), the lifecycle of the data (from generation to final destruction), the responsibilities of each party, who can see the data, what the data can and can not be used for, how the data should be protected and what should be done in the case of a security incident with the data. Credit and attribution (such as acknowledging providers in published work) is something which needs to be agreed upon and made part of a formal agreement. However, it should be noted that it may be that the provider does not wish to be acknowledged.

Trust was seen as something which could be developed through reputation. Be this through formal accreditation or certification. Data providers saw value in terms of certification such as ISO27001 \footnote{https://www.iso.org/isoiec-27001-information-security.html} used by organisations which handle secure data. Likewise, evidence of resilience to PEN testing was also seen as of value. Both data providers and consumers highlighted existing security clearance systems, often run by government agencies, as a way of highlighting trust-building. 

Patience, on the side of the data consumer, was seen as a valued attribute. Along with the ability to appreciate the `cost' (in terms of effort, value and potential risk) for the other party. 

The concept of provenance was seen as something which could help the consumer achieve a greater level of trust from the provider. In essence, being able to say where their data had gone, how it had been used and by whom would help convince the provider that the consumer could be trusted. It was also noted that this could be used in the case of a data breach to determine the loss and the potential impact.

\subsection{Who should be involved?}

It was seen from both producers and consumers that an interdisciplinary team was good for best practice. This allows for more than just a single viewpoint to be taken -- such as a computer scientist only wanting to produce an AI approach which can solve a problem without evaluating the other implications.

\subsection{How best to interact}

Two models for how data consumers and data providers interact emerged from the event. Those of the one-to-one interaction and the Trusted Third Party (TTP) which could allow for one-to-many and even many-to-many interactions. 

The one-to-one interaction was the only interaction type which consumers and providers had direct experience of. The approach was seen as providing the ability for building up a close relationship between consumer and producer -- often over a long period of time where things started from simple cases with low-security data and, as trust was built, moving up to more substantial and high-security data. The main drawbacks were the time to develop the relationship (often much longer than a standard funding cycle) and the fact that you could only make use of the relationships that you already had.

Multiple suggestions emerged for the Trusted Third Party (TTP) approach. This would allow a separate entity to act as holder and securer of the data. If the TTP is trusted by all then they can take ownership of the data, it’s security and sharing. Relationships with the TTP could be made by many people which would allow one-to-many or even many-to-many data sharing activities. However, this was seen to have the dis-advantage that the mutual understanding and trust built up between particular consumers and producers would not be present. This may diminish the chances that data providers would be willing to contribute data. 

TTP could provide some level of vetting and certification for consumers. This could enable providers the opportunity to allow (or deny) access to the data based on this, or the role could be delegated to the TTP. The funding model for a TTP was seen as a big problem. Solutions could include consumers paying to use the data or providers paying for solutions to their problems. 

It was suggested that the TTP could take on many of the tasks for the data such as anonymisation and curation. However, there was a concern that the provider would require a very high degree of trust with the TTP in order to give complete data to them. 

In order for a TTP to work it was envisaged that there would need to be a solid legal framework behind it. This would entail standard agreements for both data providers and data consumers who wished to take part. Clear definitions would also be needed in order to define who has responsibility when things go wrong.

\subsection{Communication}

There was much discussion on the issues of communication between different parties. Primarily in the context of communication between academia and other parties -- with academics being viewed as naive by industry for their simplistic view on how `the real-world works'. But this can also be present between (or within) any of the possible parties. This can be compounded due to international and cultural differences. Proposed solutions included not assuming anything and always asking questions (at all levels), avoidance of the use of colloquial terms and acronyms used within domains. Misunderstandings should be expected and effort should be made to identify these early on -- perhaps through the use of glossaries and/or ontologies.

\subsection{What data should be shared and how should data be shared?}

The exact nature of the data to be shared cannot be universally defined and would be the subject of the agreement between the provider and consumer. Here instead we discuss those general outlines of what should be shared. The quality of the data should be high -- fit for purpose, captured using high-quality and reliable methods. Where labels on the data are to be provided these should be again of high quality and as reliable as possible. Noise in the data should be kept to a minimum or at least quantified. 

There is a desire -- at least on the part of the data consumer -- that data should be provided in common formats. However, it was appreciated that providers may not be able to provide data in these formats and it was appreciated that often the consumer would need to do the work to convert to standard formats. In either case what the data is should be clearly defined. Defining how the data was captured, what each element within the data is, along with ranges of valid values. Metadata should be provided where possible along with data schemas. A decision should be made as to whether the consumer has access to a live data stream or only historical `dumps' of the data.

It was mentioned that not all data was of the same level of sensitivity and as such this should be taken into account when preparing and exchanging the data, such as anonymising it. Anonymisation was seen by all as a key requirement when discussing how data would be shared. Three levels of anonymisation were identified:

\begin{itemize}
	\item Full anonymisation - in which any highly sensitive data (such as personally identifiable information) would be removed completely from the dataset. This, however, can be very restrictive as it may remove data which is required in order to develop AI to predict what is required.
	\item Medium anonymisation - in which any highly sensitive data is replaced with a hash of that data. This minimises the chance of de-anonymisation, however, it still allows for AI techniques to be developed using the hashed data. Care needs to be taken in the choice of hashing approach relative to what the original data in order to allow the hashed value to be used as a proxy for the true data.
	\item Low anonymisation - in which any highly sensitive data is encrypted. Again, this allows for AI techniques to be developed, but it also allows the owner of the encrypted data to decrypt the data to identify what the original data was. 
\end{itemize}

Who should perform the anonymisation needs to be considered. However, the consensus from the workshop was that the provider should do this.

In general, the principle of least access -- only providing the necessary data -- was seen as best practice. However, it was appreciated that at the outset it may not be understood what the necessary data is. The volume of data needed for the work is also an important issue. This needs to weigh up the benefits that greater data volumes will have for the consumer against the `cost' for the provider in generating the data, both in time and money. 

How the data is shared needs to be clearly defined. Is the consumer allowed to store the data locally? Is the data shared through an online mechanism or an offline mechanism (for example a USB thumb drive)? If online what are the access controls? Is the data encrypted during the sharing process? In all cases encryption was seen as essential. 

\subsection{Discussion}

In summary, at the heart of the data sharing problem are three sets of issues to be considered before, during and after the data acquisition process. Before acquisition, consumers have to be conscious of the fact that they have no absolute right to data, so access to it is at the discretion of the provider. Plus, the process of accessing the data is very time consuming because of existing protocols. Very often, consumers are not sure about what the data set looks like or what data is held, so it is often the case that they are not yet clear about their own outcomes -- there is an interregnum between understanding what is available and stating what data is needed. So, consumers sometimes find it hard to be clear about their data requirements, especially the case when definitions of data can differ. Some data is pure victim content, whereas other data may be related to system data, such as logs. It is therefore very important to share research aims with the data providers. Also, it is important to establish whether the data they hold exists, and also that they own it and are in a position to share it? Is it legally (\eg GDPR) compliant? Think about why the data provider should share their data, what do they get from it? Has the consumer offered to give them an analysis?

Very often delays in access can be caused by `the human problem' whereby the fact that senior management says that they will give  access does not necessarily mean you will get the data. Very often someone (an employee down the management line) will have to spend time extracting the data, giving them additional work. Or they may be worried that they have not been collecting the data correctly. Such human impediments can delay or even thwart access. They are often expressed in very detailed data processing agreements. 

The issues before acquisition differ to those occurring during the acquisition process. Consumers will need to evolve the relationship to develop trust. Start with a memorandum of understanding (MOU) and allow this to develop as the relationship and trust develops. It should also include how to resolve disagreements about inference from data. Also, develop a formal agreement, and consider how to sustain the relationship by maintaining expectations. Remember that the original contacts will move on and new ones come in, keep them on script. Finally set up a mechanism to keep the data owner updated about the findings. 

After the acquisition process has been completed, the consumer should keep to the agreement so it will not come back to bite them, especially on keeping and deleting data. Likewise the consumer should be clear about the right to publish, they may have the data but not own it. This needs to be included in the agreement. Also, the consumer needs to be clear to acknowledge sources and even share accreditation (where the provider agrees with this); `you have nothing to lose and everything to gain'. Finally, the consumer should keep in touch with the provider as this will help in data analysis. 

\section{The way forward and beyond the cybersecurity data sharing paradox}
\label{sec:forward}

The aim of this position paper is to stimulate debates so that we can collect the right data for the job, make sure it has integrity (\eg not contaminated and fit for purpose), and to help build up trust in the data collected to enable the subsequent analysis to be trusted. Most examples of data sharing tend to be one-to-one relationships. Sometimes they are disguised as partnerships, even multiple partnerships and even collaborative examples, but upon examination they are often one-way conversations or dominated by a major organisation or business who has an interest in the outcome. This statement, whilst not referenced, is based upon a colloquial observation of a small number of `data partnerships'. 

As stated in the introduction, we need to agree as to what the central problem is, for example, cybercrime, and also agreement about what the end goal is, to prevent and mitigate its effects and prosecute offenders. Because of the various combinations of relationships outlined earlier, it is probably more practical to suggest a cross-sector (divided) solution that involves establishing sector-based and co-owned third party organisations that can take data from the partners, anonymise and aggregate it and then share with others, perhaps via another layer of organisation. This would formalise relationships, whilst also meaning that time-consuming relationship building, now depended upon, would not need to be formed, for example between academia and law enforcement or cybersecurity.  

\begin{figure}[t!]
	\centering
	\includegraphics[width=0.99\linewidth]{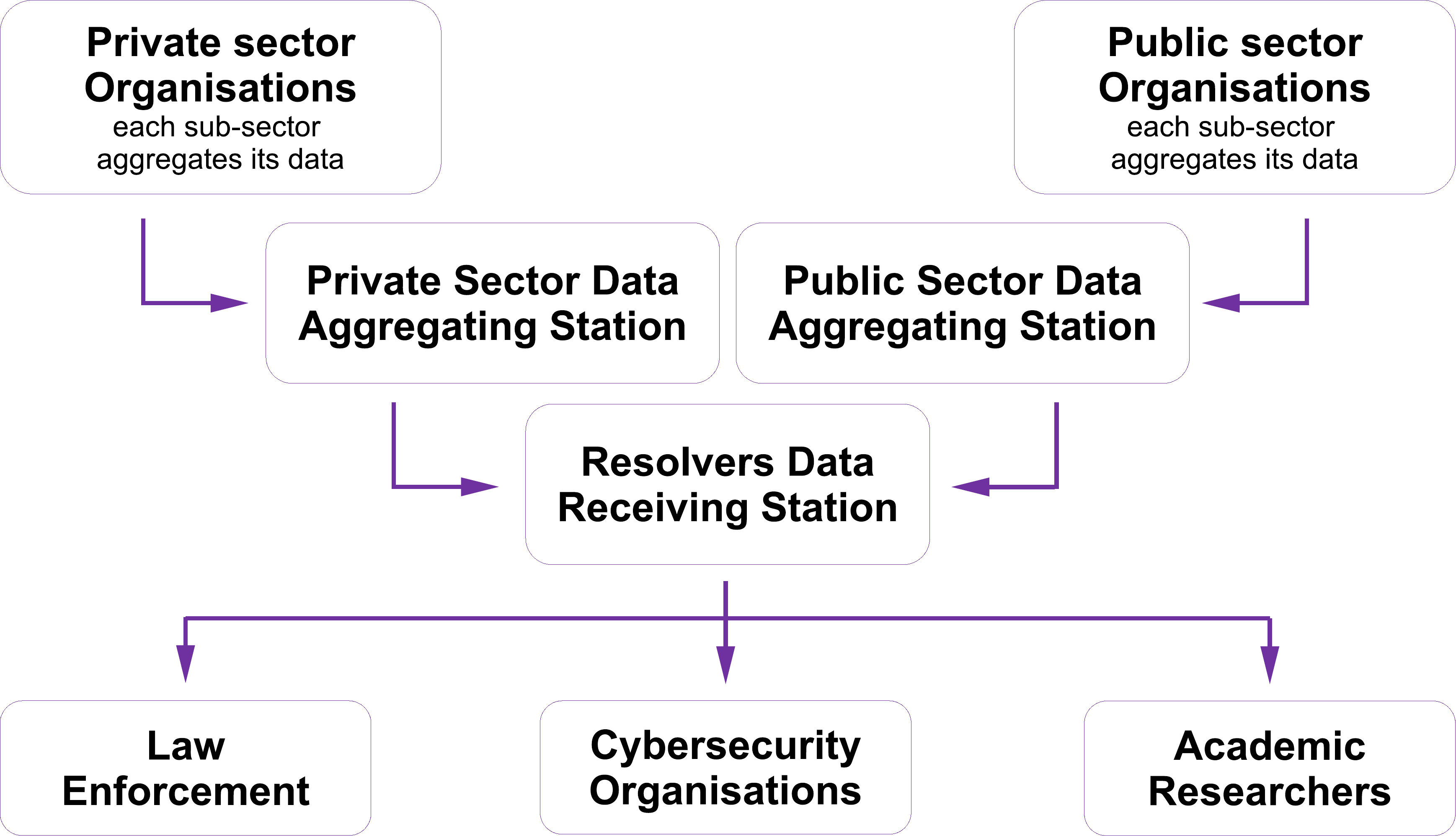}
	\captionsetup[figure]{skip=7pt}
	\captionof{figure}{Network of third party owned data aggregating stations.}%
	\label{fig:networks}\vspace{-0.5cm}
\end{figure}

Whilst such a setup is not without its challenges, it does avoid the need for unnecessary relationships and also provides mutual trust via its constitution. It does, of course, require agreed common standards and best practices in order to work. 

Drawing upon existing models of success, notably the UK Payments (APACs), mentioned earlier, and the VERIS models, both third-party organisations set up and jointly owned by a specific sector to which members submit data. The organisation then anonymises and aggregates in order to use it on behalf of the sector, but the aggregated data can then be shared with other sectors. A potential pipeline for how this could be formulated is presented in Figure \ref{fig:networks}, where data from different sectors is handled independently before being aggregated and anonymised before being made available to interested parties. At each level appropriate security and provenance approaches can be used to maximise the trust built within and about the system.

Such a proposal is not going to be simple to achieve and it is not going to be cheap and will require buy in from complete sectors and it is going to require some out of the box thinking. But the potential benefits in terms of sharing and using data for security purposes and also a range of other uses is considerable. 

So, how do we develop standards for data collection so that it can be aggregated and who do we standardise them with? Is this, for example, a potential space for the application of AI to help analyse and possibly make connections between points on the collected data? By mentioning AI in cybersecurity, it is important to balance expectations of AI and not allow claims to exceed what can be delivered. Also, to make sure that the cybersecurity solutions created are blended (like the cybercrime problem itself) and more sophisticated the current `whack-a-mole' approach, whilst also bearing in mind that AI skill sets are different from traditional science thinking in terms of, say, replicability. For example, when testing a system, running the same routine twice may produce different results is not in the Popperian mould. AI solutions should focus upon specific problems and be science and social science led.

\subsection{The Practicalities of Data Collection and Sharing}

Even when all the incentives and infrastructures required for an effective data sharing system are in place, there are certain practical considerations that need to be considered before the data is collected and curated for sharing. The data to be shared needs to be safe and reliable and should serve a specific objective before it can be shared. Of course, the required characteristics of the collected data highly depend on the nature and the purpose of the data.

For instance, a security-related data collection system needs to adhere to certain functional and security requirements \cite{lin2018survey}. In the following, we list some of the attributes that need to present in any data collection system:

\begin{itemize}
	\item The system needs to know when and from where to collect the data\cite{waku2015robust}.
	\item The system should be capable of dynamically loading information about which data to collect and storing collected data on a storage device\cite{waku2015robust}.
	\item The system must be able to export the data to other systems or external databases\cite{waku2015robust}.
	\item The system should be capable of managing and controlling the data during the collection process\cite{nouali2015ble}.
	\item The system should be efficient and stable, not interfere with the data during the collection process and should avoid computationally intensive operations \cite{ariyapala2016host, sun2004mobility}.
	\item The collection system must be flexible and scalable with respect to the amount and bandwidth of the data\cite{gad2015monitoring}.
	\item The system should be able to learn and adapt to changes in the environment where the data is generated\cite{ji2009novel}.
	\item The data collection should not introduce any noise into the environment which might affect the quality of the collected data\cite{slaviero2006active}.
	\item The data collection system should prevent any form of data loss to ensure the integrity of the collected data\cite{waku2015robust}.
	\item The system must strive to protect user privacy during the data collection process\cite{bonelli2011towards}.
	\item The system should be capable of preventing any data leakage and verify the integrity and authenticity of the collected data\cite{lin2018survey, waku2015robust}.
\end{itemize}

\section{Conclusions and future work: the need to scale up cybersecurity}
\label{sec:conclusion}

We have in this paper set out the case for engaging in a discussion (and outlining questions) about the nature of, and potential solutions for, the cybersecurity data sharing paradox in order to scale up cybersecurity by using a co-production approach towards data sharing. We have addressed the key principles which need to be addressed and we have also made some suggestions about how we can take them forward. These suggestions are intended to practically embrace the micro-politics of the world in which research takes place and address the feasibility of progressing from principles to practice in order to maintain the integrity of the data. 

In the cybersecurity arms race that is constantly evolving with offenders, we need to not only learn from their cybercrime playbook, but also be in a position to develop (and respond) with AI, derived from the data, that is one step ahead. In order to do this, we need to identify good quality and appropriate data for the application, but also agree about common standards which can be applied to data collection.

In so doing, we will need to get rid of cultural obstacles to break down siloed thinking to get ``Security through knowledge rather than obscurity''.

Finally, it is crucial that we work towards developing partnerships that co-own the problem in order to co-produce the solution. Cybercrime is not going to go away, and as society becomes more digitised and networked, then the stakes will become even higher.

\section*{Acknowledgement}

This work was in part supported by the EPSRC EMPHASIS (EP/P01187X/1), CRITiCaL (EP/M020576/1) projects and supported through the Alan Turing Institute.

\bibliographystyle{IEEEtran}
\bibliography{paper_ref}

\end{document}